\documentclass[a4paper, 11pt]{article}
\usepackage[a4paper, margin = 3cm]{geometry}

\usepackage[style=numeric-comp, url=false, doi=true, isbn=false, sorting=none, maxcitenames=2, date=year, maxbibnames=30, giveninits=true, autocite = superscript]{biblatex}

\usepackage[colorlinks=true, citecolor=blue, linkcolor=black, urlcolor=blue]{hyperref}
\usepackage{booktabs}
\usepackage{lineno}
\usepackage{authblk}
\usepackage[autostyle]{csquotes}  
\usepackage{amsmath}
\usepackage{mathtools}
\usepackage{algorithm}
\usepackage{algpseudocode}
\usepackage{amssymb}
\usepackage{stmaryrd} 
\usepackage{bm}
\usepackage{breqn}
\usepackage{enumitem}
\usepackage[font=footnotesize]{caption}
\usepackage{graphicx, wrapfig, rotating}
\usepackage{xr}
\usepackage[edges]{forest}
\algnewcommand\algorithmicforeach{\textbf{for each}}
\algdef{S}[FOR]{ForEach}[1]{\algorithmicforeach\ #1\ \algorithmicdo}

\renewbibmacro{in:}{}

\DeclareFieldFormat{pages}{#1}

\providecommand{\keywords}[1]{\textbf{Keywords:} #1}

\begin{document}

\pagenumbering{arabic}
\date{}
\title{Discrepancy measures for sensitivity analysis}

\author[1]{Arnald Puy\thanks{Corresponding author}}
\author[2]{Pamphile T. Roy}
\author[3]{Andrea Saltelli}

\affil[1]{\footnotesize{\textit{School of Geography, Earth and Environmental Sciences, University of Birmingham, Birmingham B15 2TT, United Kingdom. E-mail: \href{mailto:a.puy@bham.ac.uk}{a.puy@bham.ac.uk}}}}

\affil[2]{\footnotesize{\textit{Quansight, Vienna, Austria.}}}

\affil[3]{\footnotesize{\textit{Barcelona School of Management, Pompeu Fabra University, Carrer de Balmes 132, 08008 Barcelona, Spain.}}}

%\linenumbers

\maketitle

\begin{abstract}

While sensitivity analysis improves the transparency and reliability of mathematical models, its uptake by modelers is still scarce. This is partially explained by its technical requirements, which may be hard to understand and implement by the non-specialist. Here we propose a sensitivity analysis approach based on the concept of discrepancy that is as easy to understand as the visual inspection of input-output scatterplots. Firstly, we show that some discrepancy measures are able to rank the most influential parameters of a model almost as accurately as the variance-based total sensitivity index. We then introduce an ersatz-discrepancy whose performance as a sensitivity measure matches that of the best-performing discrepancy algorithms, is simple to implement, easier to interpret and orders of magnitude faster.
 
\end{abstract}

\newpage

\section{Introduction}

Sensitivity Analysis (SA) studies how the uncertainty in a model output can be apportioned to uncertainty in the model inputs~\cite{Saltelli2002d}. Modelers and system analysts use SA to explore how a range of different assumptions on the model parameters, boundary conditions or hypotheses affect a system of interest; to identify influential and non/influential inputs to guide further research and/or to simplify the dimensionality of the problem at hand; and to inform decision-making~\cite{razavi2021}. Due to its capacity to pry the model black box open and increase the transparency of the modeling process, SA is a cornerstone of responsible modeling~\cite{saltelli2020a, saltelli2023d}. It is also a recommended practice in the modeling guidelines of the European Commission, the Intergovernmental Panel on Climate Change or the US Environmental Protection Agency~\cite{ europeancommission2021, EPA2009}.

After more than 50 years of development, modelers dispose of several SA procedures and of a rich literature informing on which methods are most efficient in each specific SA setting~\cite{puy2022f, becker2020}. We briefly mention here some of these routines without further description and direct the reader to existing references:

\begin{itemize}[noitemsep]
\item Variance-based methods: they are well embedded in statistical theory (ANOVA), can treat sets of factors and can be used in problem settings such as ``factor fixing'' (which factor/s are the least influential and hence can be fixed to simplify the model) or ``factor prioritization'' (which factors convey the most uncertainty to the model output)~\cite{saltelli2008}.

\item Moment-independent methods: they assess sensitivities based on the entire probability distribution of the model output and may be preferred over variance-based ones when the output is long-tailed~\cite{Borgonovo2016b}.

\item The Variogram Analysis of Response Surfaces (VARS): it computes sensitivities using variograms (functions to characterize the spatial covariance structure of a stochastic process) and is especially suited to inform about the local structure of the output~\cite{Razavi2016b, Puy2021a}.

\item Shapley coefficients: it relies on Shapley values (the average marginal contribution of a given feature across all possible feature combinations). It is a good alternative to the SA approaches just mentioned when model inputs are correlated~\cite{Owen2014, song2016}.

\end{itemize}

Despite this abundance of methods, there is still a scarce uptake of SA in mathematical modeling. When performed, it is often conducted by moving ``One variable-At-a-Time'' (OAT) to determine its influence on the output, an approach that only works in low-dimensional, linear models~\cite{saltelli2010b}. A key reason behind this neglect is that proper SA methods are grounded on statistical theory and may be hard to understand and implement by non-specialists~\cite{saltelli2019a}.

Here we propose an SA measure whose use and interpretation requires little to no statistical training and that is as intuitive as the visual inspection of input-output scatterplots. Since the presence (absence)  of ``shape'' in a scatterplot indicates an influential (non-influential) input, we build on the concept of discrepancy (the deviation of the distribution of points in a multi-dimensional space from the uniform distribution) to turn discrepancy into a sensitivity measure. We show that some discrepancy algorithms nicely match the behavior of the total-order sensitivity index, a variance-based measure which estimates the first-order effect of a given input plus its interactions with all the rest~\cite{homma1996}. We also present a simple-to-implement "ersatz" discrepancy whose behavior as a sensitivity index approximates that of the best-performing discrepancy algorithms at a much more affordable computational cost. Our contribution thus provides modelers with a straightforward SA tool by turning the concept of discrepancy upside down --from a tool to inspect the input space of a sample to an index to examine its output space.  

\section{Materials and methods}

\subsection{The link between scatterplots and discrepancy}
\label{sec:link}

\begin{figure}[ht]
\centering
\includegraphics[keepaspectratio, width=\textwidth]{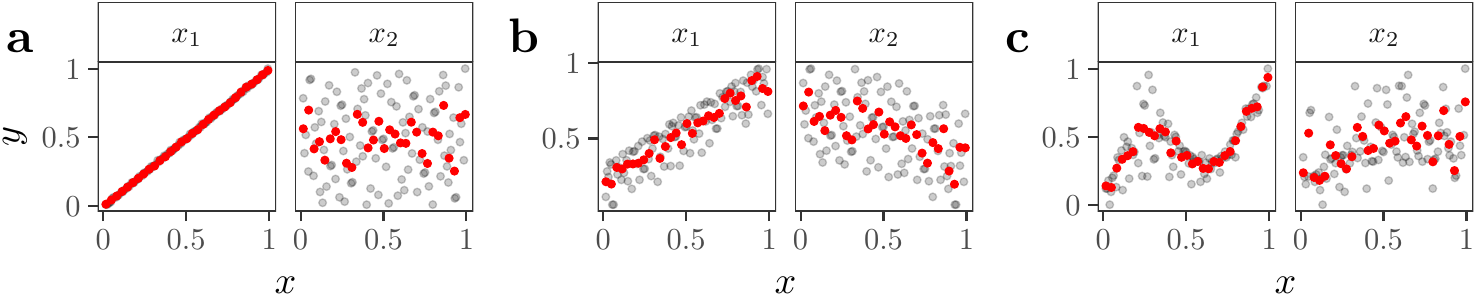}
\caption{Scatterplots of $x_k, k=1,2$ against $y$ for three different two-dimensional functions. The red dots show the running mean across 100 simulations. The functions are F1, F2 and F3 in Azzini and Rosati~\cite{azzini2022}. In a), $y$ is completely driven by $x_1$ while $x_2$ is non-influential. In b), $x_1$ is more influential than $x_2$ given its sharper trend. In c), $x_1$ is more influential than $x_2$ given the presence of larger areas where points are more rarefied.}
\label{fig:scatter}
\end{figure}

Due to their ease of interpretation, scatterplots are widely used in SA as a preliminary exploration of sensitivities before embarking on more quantitative approaches~\cite{pianosi2016}. To understand the rationale, let us first define $\bm{\Omega} = [0, 1)^d$ as a $d$-dimensional unit hypercube formed by $N_s$ sampling points and represented by the matrix $\mathbf{X}$, such that
 
\begin{equation}
\mathbf{X} = 
\begin{bmatrix}
x_1^{(1)} & x_2^{(1)} & \cdots & x_d^{(1)} \\
x_1^{(2)} & x_2^{(2)} & \cdots & x_d^{(2)} \\
\cdots & \cdots & \cdots & \cdots \\
x_1^{(N_s)} & x_2^{(N_s)} & \cdots & x_d^{(N_s)} 
\end{bmatrix} \\,
\end{equation}
where $x_k^{(i)}$ is the value taken by the $k$-th input in the $i$-th row, and $\bm{x}^{(i)}=(x_1^{(i)},\hdots,x_d^{(i)})$. Let 

\begin{equation}
\bm{y} = 
\begin{bmatrix}
y^{(1)} \\
y^{(2)} \\
\hdots \\
y^{(N_s)} \\
\end{bmatrix}
=
\begin{bmatrix}
f(\bm{x}^{(1)}) \\
f(\bm{x}^{(2)}) \\
\hdots \\
f(\bm{x}^{(N_s)}) \\
\end{bmatrix}
\end{equation}
be the vector of responses after evaluating the function (model) $f(.)$ in each of the $N_s$ rows in $\mathbf{X}$. If $y$ is sensitive to changes in $x_k$, a scatterplot of $\bm{y}$ against $\bm{x}_k$ will display a trend or shape, meaning that the distribution of $y$-points over the abscissa (over input $x_k$) will be non-uniform~\cite[28]{saltelli2008}. Generally, the sharper the trend/shape, the larger the area without points and the stronger the influence of $\bm{x}_k$ on $\bm{y}$. In contrast, a scatterplot where the dots are uniformly distributed across the space formed by $\bm{x}_k$ and $\bm{y}$ evidences a totally non-influential parameter~(Fig.~\ref{fig:scatter}a--c). This heuristic suggests that, in a 2-dimensional space, the deviation of points from the uniform distribution can inform on the extent to which $\bm{y}$ is sensitive to $\bm{x}_k$.

There are different ways to assess the ``uniformity'' of a sample. Geometrical criteria such as \textit{maximin} or \textit{minimax} respectively maximize the minimal distance between all points or minimize the maximal distance between any location in space and all points of the sample~\cite{pronzato2017a}. These criteria are notably used in circle packing problems. In contrast, uniformity criteria measure how the spread of points deviates from a uniform spread of points (in the sense of a multi-dimensional uniform distribution): taking a subspace of the parameter space $J_x =[0, x)$, we count the number of points $N_{s_{[0, x)}}$ in the subspace and compare it to the total number of the points $N_s$ of the sample. The resulting value is subtracted by the volume of the subspace $\textrm{Vol}_{[0, x)}$, 

\begin{align}
\left\vert \frac{N_{s_{[0, \bm{x})}}}{N_s} - \textrm{Vol}_{[0, \bm{x})} \right\vert.
\label{eq:uniformity_criteria}
\end{align}

The resulting quantity is known as the discrepancy at point $x$. Notice that with this description, the origin of the domain ($[0]^d$) is part of every subspace.

Several measures calculate the discrepancy over the whole domain, assumed to be the unit hypercube. The $L_p$-discrepancy measure, for instance, takes the average of all discrepancies, as 

\begin{align}
D_p = \left\{ \int \left\vert \frac{N_{s_{[0, x)}}}{N_s} - \textrm{Vol}_{[0, x)} \right\vert^p\,dx \right\}^{1/p}.
\label{eq:L_p_discrepancy}
\end{align}

When $p\rightarrow\infty$, the measure is known as the ``star discrepancy'', which corresponds to the Kolmogorov-Smirnov goodness-of-fit statistic~\cite{fang2018}.

\begin{align}
D^* = \sup_{x \in \mathbf{X}} \left\vert \frac{N_{s_{[0, x)}}}{n} - \textrm{Vol}_{[0, x)} \right\vert.
\label{eq:L_inf_discrepancy}
\end{align}

When $p\rightarrow2$, the measure is known as the ``star $L_2$ discrepancy''~\cite{Warnock1972}, which corresponds to the Cramér-Von Mises goodness-of-fit statistic~\cite{fang2018}. Its analytical formulation reads as 

\begin{equation}
SL^2(\mathbf{X}_d^{N_s}) = 3^{-d} - \frac{2^{1-d}}{N_s} \sum_{i=1}^{N_s} \prod_{k=1}^d \left [ 1 - (x^{(i)}_k)^2 \right ] + \frac{1}{{N_s}^2} \sum_{i=1}^{N_s} \sum_{j=1}^{N_s} \prod_{k=1}^d \left [1 - \max(x^{(i)}_k, x^{(j)}_k) \right ]\,.
\label{eq:L_p_star}
\end{equation}

The ``modified discrepancy'' $M^2$ slightly varies the ``star $L_2$ discrepancy''~\cite{Franco2008}, and reads as

\begin{equation}
M^2(\mathbf{X}_d^{N_s}) = \left ( \frac{4}{3} \right ) ^d - \frac{2^{1-d}}{N_s} \sum_{i=1}^{N_s} \prod_{k=1}^d \left [ 3 - (x^{(i)}_k)^2 \right ] + \frac{1}{{N_s}^2} \sum_{i=1}^{N_s} \sum_{j=1}^{N_s} \prod_{k=1}^d \left [2 - \max (x^{(i)}_k, x^{(j)}_k) \right ]\,.
\label{eq:L_p_modified}
\end{equation}

Fang et al.~\cite{Fang2006} proposed the following criteria to assess the quality of discrepancy measures:
\begin{enumerate}[noitemsep]
	\item They should be invariant under permuting factors and/or runs.
	\item They should be invariant under coordinate rotation.
	\item They should measure not only uniformity of the hypercube, but also of any sub-projections.
	\item They should have some geometric meaning.
	\item They should be easy to compute.
	\item They should satisfy the Koksma–Hlawka inequality.
	\item They should be consistent with other criteria in experimental design, such as the aforementioned distance criteria.
\end{enumerate}

Equations~\ref{eq:L_p_discrepancy}--\ref{eq:L_p_modified} do not satisfy these criteria as they lack sensitivity, vary after rotation and consider the origin to have a special role. To mitigate these issues, some modified formulations of the $L_2$-discrepancy have been proposed. As we shall see, these methods treat the corners of the hypercube differently. The ``centered discrepancy'' $C^2$, for instance, does not use the origin of the domain when selecting samples to create the volumes, but the closest corner points $J_x$~\cite{Hickernell1998}, as

\begin{align}
C^2(\mathbf{X}^{N_s}_d) =& \left( \frac{13}{12} \right)^d - \frac{2}{N_s}\displaystyle\sum_{i=1}^{N_s}\prod_{k=1}^{d} \left( 1 + \frac{1}{2} \mid  x_k^{(i)} - 0.5\mid - \frac{1}{2} \mid  x_k^{(i)} - 0.5\mid^2\right)\\ \nonumber
& + \frac{1}{N_s^2}\sum_{i,j=1}^{N_s}\prod_{k=1}^d \left( 1 + \frac{1}{2} \mid  x_k^{(i)} - 0.5\mid + \frac{1}{2} \mid  x_k^{(j)} - 0.5\mid - \frac{1}{2} \mid  x_k^{(i)} - x_k^{(j)}\mid \right).
\end{align}

The ``symmetric'' discrepancy $S^2$ is a variation of the centered discrepancy that accounts for the symmetric volume of $J_x$~\cite{Hickernell1998}:

\begin{align}
S^2(\mathbf{X}^{N_s}_d) =& \left( \frac{4}{3} \right)^d - \frac{2}{N_s}\displaystyle\sum_{i=1}^{N_s}\prod_{k=1}^{d} \left( 1 + 2 x_k^{(i)} - 2 \left(x_k^{(i)}\right)^2 \right)\\ \nonumber
& + \frac{2^d}{N_s^2}\sum_{i,j=1}^{N_s}\prod_{k=1}^d \left( 1 - \mid  x_k^{(i)} - x_k^{(j)}\mid \right).
\end{align}

The ``wrap-around discrepancy'' $WD^2$, on the other hand, does not use any corners nor the origin (hence it is also called ``unanchored discrepancy'')~\cite{Hickernell1998b}. 

\begin{equation}
WD^2(\mathbf{X}^{N_s}_d) = -\left( \frac{4}{3} \right)^d + \frac{1}{N_s^2}\displaystyle\sum_{i,j=1}^{N_s}\prod_{k=1}^{d} \left( \frac{3}{2} -\mid  x_{k}^{(i)} - x_{k}^{(j)} \mid + \mid  x_{k}^{(i)} - x_{k}^{(j)} \mid^2 \right).
\label{eq:wraparound}
\end{equation}

The centred discrepancy $C^2$ and the wrap-around discrepancy $WD^2$ are the most commonly used formulations nowadays. As per numerical complexity, these equations have a complexity of $O(N_s^{2}d)$.

The Koksma–Hlawka inequality implies that low discrepancy sequences reduce a bound on
integration error~\cite{hickernell2006}. Hence, averaging a function over samples of points with a low discrepancy would achieve a lower integration error as compared to a random sample [also called Monte Carlo (MC) sampling]. Quasi-Monte Carlo (QMC) methods are designed with that problem in mind. For well behaved functions, they typically achieve an integration error close to $O(N_s^{-1})$. There is an extensive body of literature covering methods to create a sample with a low discrepancy. Notably, the low discrepancy sequence of Sobol'~\cite{sobol1967} is one of the most widely used method and it's randomized version nearly achieves a convergence rate of $O(N_s^{-3/2})$.

\subsection{An ersatz discrepancy}
The discrepancies presented in Equations~\ref{eq:L_p_discrepancy}--\ref{eq:wraparound} are state-of-the-art measures whose understanding demands a modicum of statistical training. They are also computationally complex given their reliance on column-wise and row-wise loops, a feature that hampers their scalability to larger sample sizes and/or higher-dimensional settings. Here we propose an ersatz discrepancy that addresses these issues and builds on the link between scatterplots, discrepancy and sensitivity discussed in Section~\ref{sec:link}. 

We suggest to split the $\bm{x}_k$, $\bm{y}$ plane into a uniform grid formed by $\lceil \sqrt{N_s} \rceil \times \lceil \sqrt{N_s} \rceil$ cells, and calculate the ratio between the number of cells with points and the total number of cells (Algorithm~\ref{alg:saltelli_algo}). The resulting value thus informs on the fraction of cells that are populated by at least one point: the closer this value is to 1, the more the design approaches a uniform distribution --and the less influential $\bm{x}_k$ is. Note that our measure (S-ersatz hereafter) is very close to the definition of discrepancy, but differs from previous discrepancy measures in that larger values indicate better uniform properties. For Equations~\ref{eq:L_p_discrepancy}--\ref{eq:wraparound}, smaller values are better because they reflect the difference between the distribution of sampling points and a distribution with an equal proportion of points in each explored sub-region of the unit hypercube.

\begin{algorithm}[ht]
\caption{The S-ersatz of discrepancy.}
\label{alg:saltelli_algo}
\begin{algorithmic}[1]
\Require $\bm{x}_k, \bm{y}$ \Comment{Model input $\bm{x}_k$ and model output $\bm{y}$, both of length $N_s$.}
\State $s \gets \lceil \sqrt{N_s} \rceil$
\State $\mathbf{O} \gets$ Create a $s \times s$ zero matrix \Comment{Each element in $\mathbf{O}$ is a grid cell.}
\State $\mathbf{M} \gets$ Column-bind $\bm{x}_k$ and $\bm{y}$.
\ForEach {$k \in \mathbf{M}$}

\State $\mathbf{M}_k \gets \lceil \mathbf{M}_k \times s \rceil$
\If{$\mathbf{M}_k^{(i)} = 0$} \Comment{Round up.}
\State $\mathbf{M}_k^{(i)} \gets 1$
\EndIf
\EndFor
\State $\mathbf{O}[\mathbf{M}] \gets 1$ \Comment{Use $\mathbf{M}$ to identify which elements in $\mathbf{O}$ should be 1.}
\State S-ersatz $\gets \sum \{o \in \mathbf{O} | o = 1\} / N_s$ \Comment{Compute the proportion of grid cells with 1.}
\end{algorithmic}
\end{algorithm}

\begin{figure}
\centering
\includegraphics[keepaspectratio, width=\textwidth]{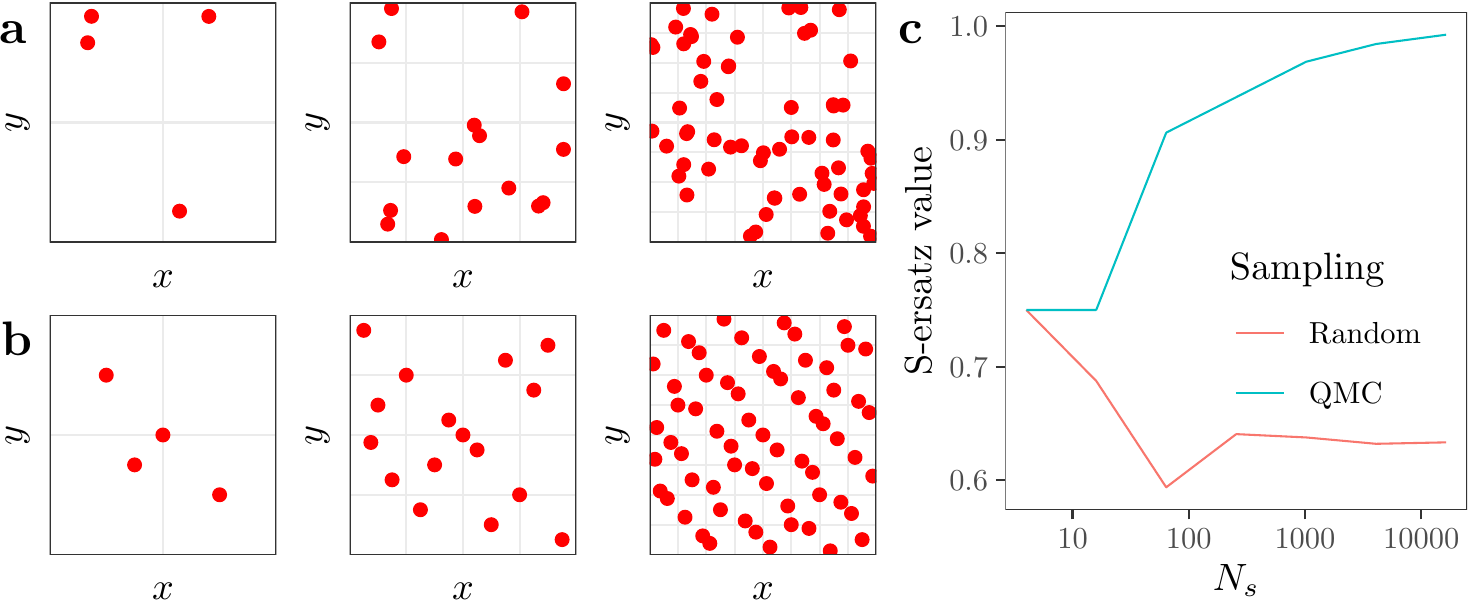}
\caption{The S-ersatz  discrepancy. a) and b) show planes with sampling points produced using random and QMC (Sobol' sequence)  respectively, for a total sample size of $N_s=2^b, b=2,4,6$. c) Evolution of the S-ersatz through different $N_s$.}
\label{fig:grid_plots}
\end{figure}

We graphically represent our approach in Figure~\ref{fig:grid_plots} for both random (a) and quasi-random (b) numbers [Quasi-Monte Carlo (QMC), using the low discrepancy sequence of Sobol']. The latter are known to outperform the former in sampling the unit hypercube by leaving smaller unexplored volumes. This means that a design with QMC should display larger S-ersatz values. For $N_s=2^2$ (first column), the plane is partitioned into four cells given that $\lceil \sqrt{N_s} \rceil \times \lceil \sqrt{N_s} \rceil = 4$. For $N_s=2^b, b=4,6$ (second and third columns), the plane is partitioned into 16 and 64 cells respectively. The ratio of sampled cells to the total number of cells is $3/4=0.75$, $11/16=0.68$ and $40/64=0.62$ in a), and $3/4=0.75$, $12/16=0.75$ and $59/64=0.92$ in b). The behavior of the S-ersatz therefore nicely matches the well-known capacity of quasi-random sequences in covering the domain of interest more evenly and quicker than random numbers (Figure~\ref{fig:grid_plots}c).

The S-ersatz is very close to the definition of the discrepancy. Instead of counting the number of points in each cell and dividing by the number of samples, it counts 1 when a cell contains a point and 0 if not. Another difference with previous discrepancy measures is that cells are pre-defined as a uniform grid. This setting corresponds to the ``property A'' described by Sobol'~\cite{Sobol1976}, which is another way of assessing the uniformity of a sample. This property is key in the design of a class of QMC methods such as Sobol's low discrepancy sequence.

The S-ersatz therefore displays the following properties:

\begin{enumerate}[noitemsep]
    \item It is an unanchored discrepancy.
    \item It is invariant under rotation and ordering.
    \item It has a geometrical meaning.
    \item It is simple to approach and implement.
    \item It is fast to compute.
\end{enumerate}

Furthermore, as opposed to classical discrepancy measures, both the upper and lower bounds are exactly known. The worst case would place all points in a single cell, while the ideal case would place a point per cell. This implies S-ersatz $\in [1/N_s, 1]$. Other methods don't have simple expressions for their lower bounds which work for all combinations of number of samples and dimensions.

\subsection{A sensitivity analysis setting}
\label{sec:SA_setting}

To explore the extent to which the concept of discrepancy is apt to distinguish influential from non-influential inputs, we benchmark the performance of Equations~\ref{eq:L_p_star}--\ref{eq:wraparound} and of the S-ersatz (Algorithm~\ref{alg:saltelli_algo}) in a sensitivity analysis setting. Specifically, we assess how well each discrepancy measure ranks the most influential parameters, i.e., those that convey the most uncertainty to the model output. Since only a small fraction of the model inputs tends to be responsible for variations in the output (a manifestation of the Pareto principle, where 80\% of the effects are caused by 20\% of the inputs~\cite{pareto1906, box1986}), most practitioners are content with the correct identification of top ranks only~\cite{sheikholeslami2019}. As a yardstick of quality, we compare the ranking produced by each discrepancy measure against the ranking produced by the Jansen estimator~\cite{jansen1999}, one of the most accurate variance-based total-order estimators~\cite{saltelli2010b, puy2022f}. We consider the total-order index to be an appropriate standard because it captures first-order as well as interaction effects~\cite{homma1996}. 

\begin{figure}[ht]
\begin{forest}
  for tree={
    grow'= 0,
    outer sep = 1pt,
    parent anchor=children,
    s sep = 0pt,
    child anchor=parent,
    anchor=parent,
    if n children=0{folder}{},
    edge path'={(!u.parent anchor) -- ++(2pt,0) |- (.child anchor)},
  },
  where n=1{
    calign with current edge
  }{},
   [\textbf{Factor} 
        [{Sampling method: $\tau\sim\mathcal{DU}(1, 2)$}
         [{1: Random numbers. } ]
         [{2: Quasi-random numbers \cite{sobol1967, Sobol1976}. } ]
         ]
        [{Sample size: $N_s\sim\mathcal{DU}(10, 100)$}]
        [{Model dimensionality: $d\sim\mathcal{DU}(3,50)$}.]
        [{Form of the test function: $\epsilon\sim\mathcal{DU}(1,200)$}.]
       [{Distribution of the model inputs: $\phi\sim\mathcal{DU}(1,8)$}.
        [{1: $\mathcal{U}(0, 1)$ } ]
        [{2: $\mathcal{N}(0.5, 0.15)$ } ]
        [{3: $Beta(8, 2)$ } ]
        [{4: $Beta(2, 8)$ } ]
        [{5: $Beta(2, 0.8)$ } ]
        [{6: $Beta(0.8, 2)$ } ]
        [{7: $Logitnorm(0, 3.16)$ } ]
        [{8: Random between 1--7 } ]
       ]
    ]
\end{forest}
\caption{Tree diagram with the uncertain inputs, their distributions and their levels.}
\label{fig:tree}
\end{figure}
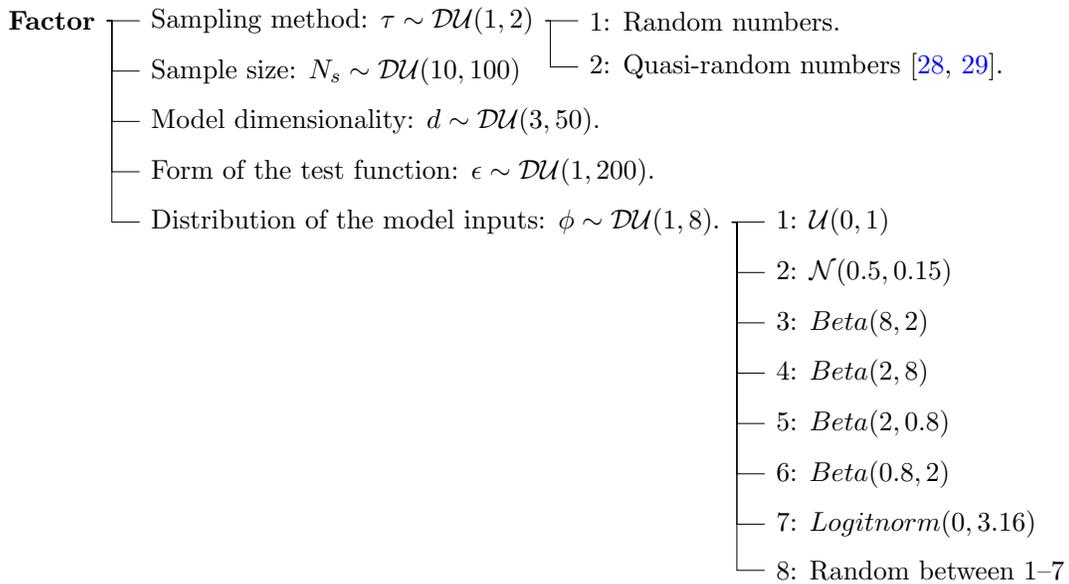

To minimize the influence of the benchmarking design on the results of the analysis, we randomize the main factors that condition the accuracy of sensitivity estimators: the sampling method $\tau$, base sample size $N_s$, model dimensionality $d$, form of the test function $\epsilon$ and distribution of model inputs $\phi$~\cite{puy2022f, becker2020}. We describe these factors with probability distributions selected to cover a wide range of sensitivity analysis settings, from low-dimensional, computationally inexpensive designs to complex, high-dimensional problems formed by inputs whose uncertainty is described by dissimilar mathematical functions (Fig.~\ref{fig:tree}). Although not exhaustive, this approach permits us to go beyond classic benchmarking exercises, which tend to focus on a handful of test functions or just move one design factor at-a-time~\cite{pianosi2015a, Azzini2021}.

We first create a $2^9 \times 5$ sample matrix using quasi-random numbers~\cite{sobol1967, Sobol1976}, where the $i$-th row represents a random combination of $\tau,N_s,d,\hdots$ values and each column is a factor whose uncertainty is described with its selected probability distribution (Fig.~\ref{fig:tree}). In the $i$-th row of this matrix we do the following:

\begin{enumerate}
\item We construct two sampling matrices with the sampling method as defined by $\tau^{(i)}$: 

\begin{itemize}
\item A $N_{s}^{(i)}(d^{(i)}+1) \times d^{(i)}$ sample matrix formed by an $\mathbf{A}$ matrix and $d^{(i)}$ $\mathbf{A}_{B_k}$ matrices, where all columns come from $\mathbf{A}$ except the $k$-th, which comes from $\mathbf{B}$. This sampling design is required to run a sensitivity analysis with the Jansen estimator~\cite{lopiano2021b, puy2022a}. We refer to this matrix as the ``Jansen matrix''.

\item A $N_{s}^{(i)}(d^{(i)}+1) \times d^{(i)}$ sample matrix formed by an $\mathbf{A}$ matrix only. Since discrepancy measures do not require a specific sampling design, we match the number of rows required by the Jansen estimator to ensure that the comparison between the latter and the discrepancy measures is done on the same total number of model runs. We refer to this matrix as the ``Discrepancy matrix''.

\end{itemize}

\item We define the distribution of each input in these matrices according to the value set by $\phi^{(i)}$ (Fig.~S1).

\item We run a metafunction rowwise through both the Jansen and the Discrepancy matrix and produce two vectors with the model output, which we refer to as $\bm{y}_J$ and $\bm{y}_D$ respectively. Our metafunction, whose functional form is defined by $\epsilon^{(i)}$, is based on the Becker metafunction~\cite{becker2020} and randomizes over 13 univariate functions representing common responses in physical systems and in classic sensitivity analysis functions (from cubic, exponential or periodic to sinusoidal, see Fig.~S2). A detailed explanation of the metafunction can be found in Becker~\cite{becker2020} and in Puy et al.~\cite{puy2022f}. 

\item We use $\bm{y}_J$ to produce a vector with the total-order indices $\bm{T}$, calculated with the Jansen estimator~\cite{jansen1999}. We also use $\bm{y}_D$ to produce seven vectors with the discrepancy values $\bm{D}$, one vector for each of the seven discrepancy measures tested.

\item We rank-transform $\bm{T}$ and $\bm{D}$ using savage scores, which emphasize and downplay top and low ranks respectively~\cite{iman1987, Savage1956}. To check how well discrepancy measures match the ranks produced with the Jansen estimator, we calculate for each discrepancy measure the Pearson correlation between $\bm{T}$ and $\bm{D}$, which we denote as $r$.

\end{enumerate}

\section{Results}

\subsection{Discrepancy measures for sensitivity analysis}
\begin{figure}[ht]
\centering
\includegraphics[keepaspectratio, width = \textwidth]{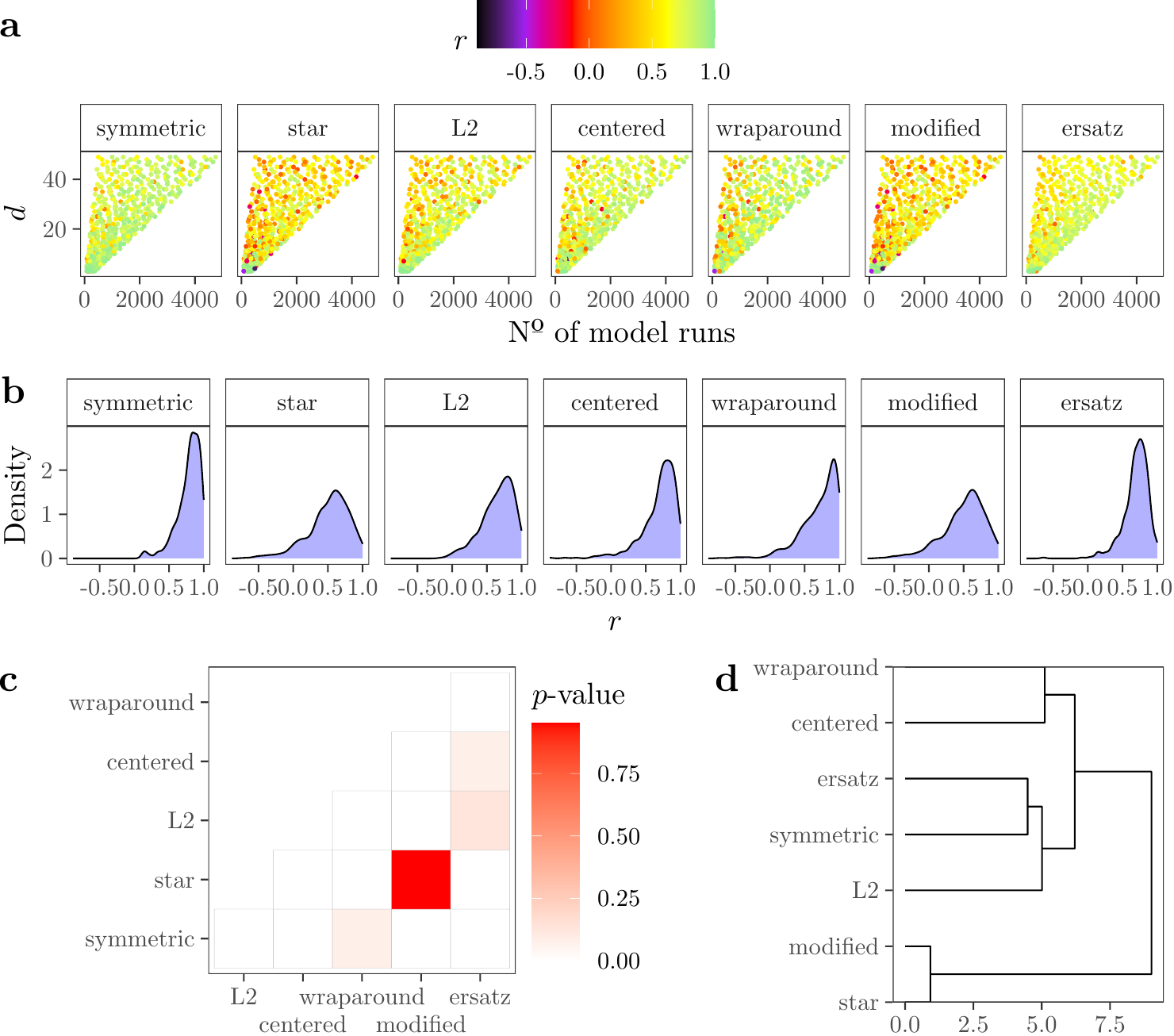}
\caption{Results of the benchmarking. a) Distribution of the Pearson correlation $r$ between the savage scores-transformed ranks yielded by each discrepancy measure and the savage scores-transformed ranks produced by the Jansen~\cite{jansen1999} estimator. Each dot is a simulation that randomizes the sample size $N_s$, the dimensionality $d$, the underlying probability distributions $\phi$, the sampling method $\tau$ and the functional form of the metafunction $\epsilon$. The total number of simulations is $2^9$. b) Density plots of $r$ values. c) Tile plot showing the $p$-values of a pairwise Mood test on medians. d) Hierarchical clustering.}
\label{fig: scatter}
\end{figure}

We conduct our analysis in a non-orthogonal domain to make the total number of model runs used by each discrepancy measure match that of the Jansen estimator, defined as $N_s (d + 1)$ (Section~\ref{sec:SA_setting}). We observe that the symmetric, the centered, the wraparound and the S-ersatz present consistently high $r$ values throughout most of the domain investigated, with lower values concentrating largely on the leftmost part of the domain (characterized by simulations with low sample sizes and increasing dimensionality). In contrast, no discernible pattern is visible for the star, L2, centered or modified measures, which comparatively present a much larger number of low and negative $r$ values (Fig.~\ref{fig: scatter}a).

Overall, the distribution of $r$ values is left-skewed for all discrepancy measures. This suggests that they are able to properly approximate the savage-transformed ranks produced by the Jansen estimator in a non-negligible number of SA settings. According to median values, the discrepancy measure that better matches the Jansen savage-transformed ranks is the symmetric ($r=0.81$), followed by the wrap-around ($r=0.78$) and the centered ($r=0.74$). The S-ersatz also displays a good performance ($r=0.72$), and its spread is as small as that of the symmetric discrepancy measure (Fig.~\ref{fig: scatter}b).

To check whether these median $r$ values come from different distributions, we conduct a pairwise Mood test on medians with corrections for multiple testing. We cannot reject the null hypothesis of a difference in medians between the wraparound and the symmetric, between the modified and the star, or between the ersatz, the L2 and the centered measure (95\% CL, Fig.~\ref{fig: scatter}c). A hierarchical cluster analysis suggests that the difference in the distribution of $r$ values is mainly between two groups: the group formed by the modified and the star discrepancy, and the group formed by all the rest. The star and the modified discrepancy present the most similar distributions, followed by the S-ersatz and the symmetric discrepancy (Fig.~\ref{fig: scatter}d). 

\begin{figure}[ht]
\centering
\includegraphics[keepaspectratio]{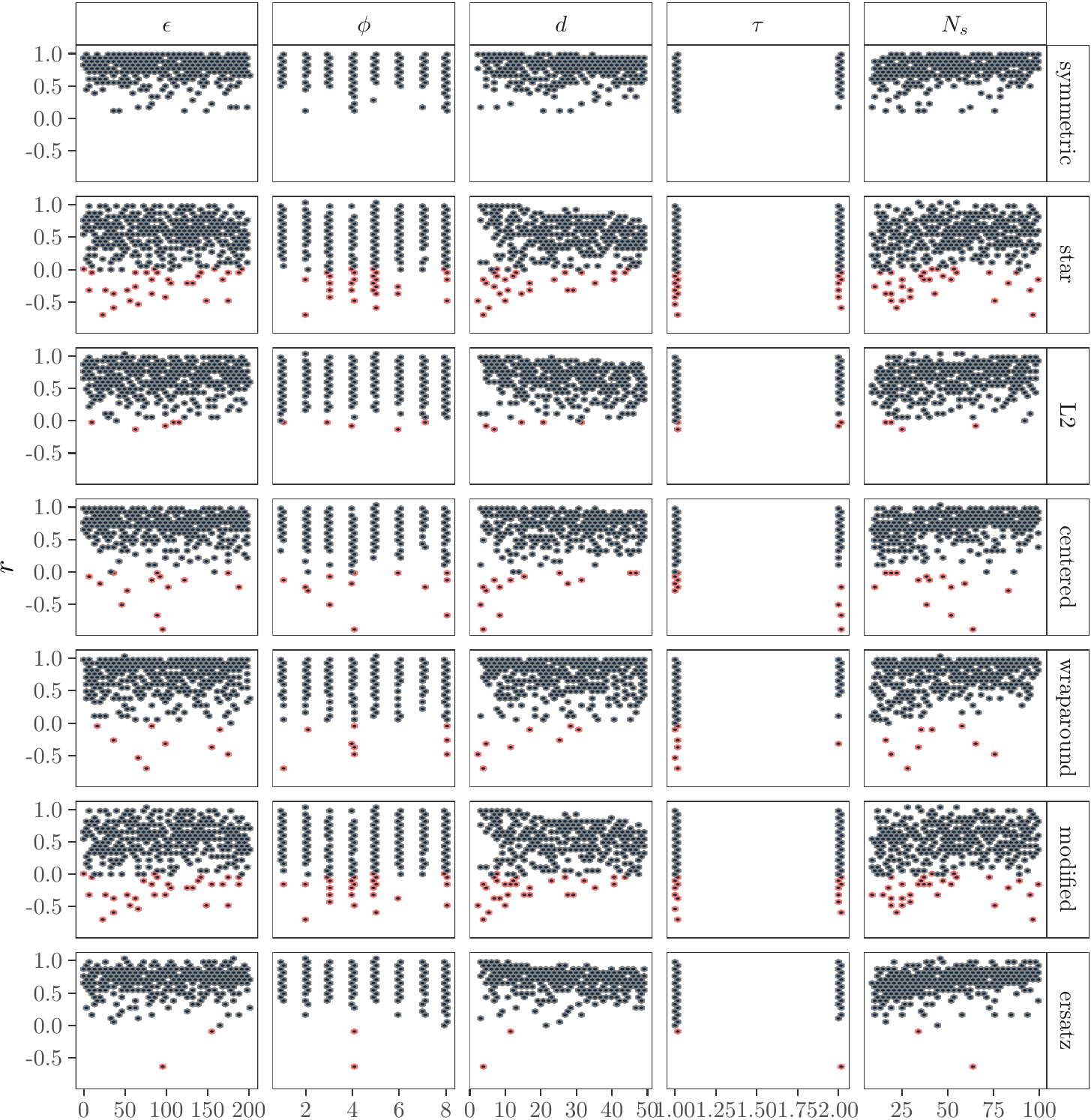}
\caption{Scatterplot of $x_k$ against the model output $r$ (the Pearson correlation) for each discrepancy measure. The red hexbins display simulations where $r < 0$. The number of simulations in each facet is $N_s=2^9$.}
\label{fig:scatterplots}
\end{figure}

The capacity of discrepancy measures in matching the savage-transformed ranks of the Jansen estimator seems to be mostly determined by high-order interactions between the benchmark factors selected in our analysis (Fig.~\ref{fig:tree}). The model dimensionality ($d$) and the base sample size ($N_s$) are the only factors with a visible direct effect on the accuracy of discrepancy measures, especially on the S-ersatz: higher dimensionalities and larger sample sizes tend to respectively diminish and increase their performance (Fig.~\ref{fig:scatterplots}). Interestingly, the variability in the performance of discrepancy measures does not seem to be critically determined by the functional form of the model ($\epsilon$), the underlying distribution ($\phi$) or the sampling method used to design the sample matrix ($\tau$).

As displayed in Figs.~\ref{fig: scatter}--\ref{fig:scatterplots}, some simulations yield $r<0$. To explore the reasons underlying this rank reversal, we plot all simulations where at least one discrepancy measure yielded $r<0$ and cross-check the other $r$ values. We observe that the production of negative $r$ values is measure-specific: in the same simulation some discrepancy measures yielded $r<0$ while others produced very high $r$ values and hence accurately matched the rankings of the Jansen estimator~(Figs.~S4--S13). This indicates that certain discrepancy measures, especially the modified and the star, may be more volatile than the rest when used in an SA setting.

\subsection{Computational complexity}

\begin{figure}[ht]
\centering
\includegraphics[keepaspectratio]{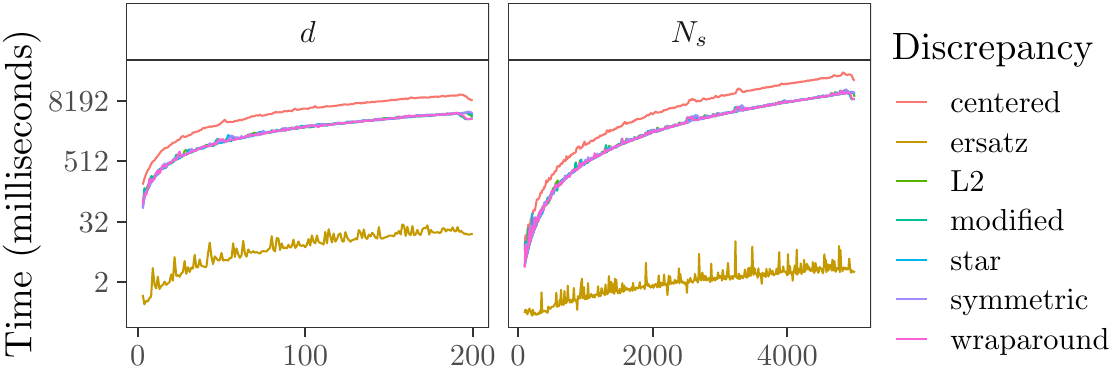}
\caption{Time complexity of discrepancy measures. In the $d$ facet $N_s=500$, while in the $N_s$ facet, $d=5$.}
\label{fig:timing}
\end{figure}

The numerical efficiency of a sensitivity analysis method (how much time it requires to run its algorithmic implementation) is an important property to take into account when deciding which SA approach to use. If the model of interest is already computationally burdensome, the extra computational strain added by a demanding SA method may make the implementation of the latter unfeasible. To pinpoint the computational requirements of discrepancy measures, we calculate the time it takes to evaluate each expression using the R package \texttt{microbenchmark}~\cite{mersmann2021}, which uses sub-millisecond accurate timing functions. We use the implementations of Equations~\ref{eq:L_p_star}--\ref{eq:wraparound} in the \texttt{R} package \texttt{sensitivity}~\cite{iooss2022}, which are written in \texttt{C++}. Our implementation of the S-ersatz algorithm uses base \texttt{R} language~\cite{Puy2022_code_discrepancy}. To gain robust insights into the time complexity of all these discrepancy measures, we explore their efficiency through a wide range of sample sizes $N_s$ (100-5,000) and dimensionalities $d$ (3-100), which we treat as random factors following the approach described in Section~\ref{sec:SA_setting}. The results, which are presented in Fig.~\ref{fig:timing}, match the expected numerical complexity of the different discrepancy measures [$O(N_s^2d)$] and of the S-ersatz [$O(N_sd)$].

\newpage

\section{Discussion and conclusions}

Sensitivity analysis (SA) is a key method to ensure the quality of model-based inferences and promote responsible modelling~\cite{saltelli2020a, saltelli2023d}. Here we contribute to the advancement of SA by proposing an approach that is as easy to understand as the visual inspection of input-output scatterplots. Our contribution is fourfold: 

\begin{enumerate}[noitemsep]
\item We turn the concept of discrepancy, usually applied to explore the input space of a mathematical simulation, into a tool to explore its output.

\item We make SA more accessible to the non-specialist by linking the sensitivity of a given input to this input's capacity to leave ``holes'' in the input-output scatterplot.

\item We provide numerical evidence that discrepancy is a good proxy of the total-order sensitivity index, a recommended best-practice in variance-based global SA~\cite{homma1996, saltelli2008}.

\item We introduce an ersatz discrepancy whose behavior in an SA setting approximates the best discrepancy measures while at a much reduced computational cost.

\end{enumerate}

It has been argued that a desirable property of a good SA method is its capacity to be translated into comprehensible language~\cite{saltelli2008}. In other words, a satisfactory SA method allows an easy explanation as to why a factor is identified as more influential than another. Variance-based methods have this property: 

\begin{itemize}[noitemsep]

\item The first-order index $S_i$ says that $x_1$ is more important than $x_2$ when fixing $x_1$ leads on average to a greater reduction in the output variance than fixing $x_2$. 

\item The total-order $T_i$ says that $x_1$ is more important than $x_2$ when fixing all factors but $x_1$ leaves on average a greater residual variance than doing the same on $x_2$. 

\item With Shapley coefficients, $x_1$ is more influential than $x_2$ if all combinations of factors including $x_1$ yields a higher outcome than the same conducted on $x_2$~\cite{Owen2014, song2016}. 

\end{itemize}

The same clarity applies when using discrepancy measures: 

\begin{itemize}
\item $x_1$ is more influential than $x_2$ when the scatterplot of $x_1$ against $y$ displays a more discernible shape than the scatterplot of $x_2$ against $y$.
\end{itemize}

Our exploration of the performance of discrepancy measures as SA tools reveals the existence of two main groups: (1) the group formed by the wraparound, the centered, the symmetric, the L2 and the S-ersatz, and (2) the group formed by the modified and the star discrepancies. The second group matches the behavior of the Jansen estimator worse than the first group. This is because both the modified and the star discrepancy give the origin of the domain ($[0]^d$) a special meaning: points further away from the origin affect less the measures. As for the wraparound, the centered, the symmetric, the L2 and the S-ersatz, their results are similar because they also treat the origin and corner points in the same way. Their differences due to the dimensionality do not play an important role here given our focus on 2D sub-projections.

Note that the quality of the sampling is paramount. As we are constructing 2D sub-projections between $x_k$ and $y$, it is important to have a uniform distribution on the sample axis. If this is not the case, the measure will be biased by a non-uniformity on the sample side. Randomized QMC methods such as Sobol' low-discrepancy sequences are appropriate as their properties guarantee the uniformity on sub-projections~\cite{sobol1967, kucherenko2015}. Essentially, it would only be necessary to ensure good 1D-subprojections, which means that simpler QMC methods such as Latin Hypercube sampling could alternatively be used~\cite{kucherenko2015}. This makes our method even more straightforward to implement in practice.

The use of discrepancy measures in an SA setting can be extended to higher dimensions to appraise high order interactions. Both the centered and the wrap-around discrepancy, however, are known to have shortcomings with regards to dimensionality: the centered suffers from the curse of dimensionality, whereas the wrap-around is not sensitive to a shift of dimension~\cite{fang2013, fang2018}. Our results may therefore change should we increase the sub-projections' dimensionality. Recently, some work has been done to combine the complementary benefits of both the centered and the wraparound discrepancy into a single measure: the mixture discrepancy $MD^2$---see~\cite{fang2013}. This method could also prove to be efficient here as it should give a more uniform importance to every part of the domain.

Finally, our work shows that ersatz methods can be a good alternative to classical discrepancy measures in an SA setting, and that there is potential to develop new ersatz discrepancies. The newly added Newcomb-Benford measure~\cite{roy2021} may be a venue worth exploring further.

\section*{Acknowledgements}
We thank Bertrand Iooss and Art Owen for their insights into discrepancy measures. All mistakes are our own. AS has worked in this paper within the framework of the i4Driving project, funded by the European Union’s Horizon Europe research and Innovation program (Grant Agreement ID 101076165). PR is partially supported by grant ``SciPy: Fundamental Tools for Biomedical Research'' (EOSS5-0000000176) from the Chan Zuckerberg Initiative DAF, an advised fund of the Silicon Valley Community Foundation.

\section*{Code availability}
The \texttt{R} code to replicate our results is available in Puy~\cite{puy2023a} and in \url{https://github.com/arnaldpuy/discrepancy}. PR will also contribute the new methods to the Python library SALib~\cite{iwanaga2022}.

% Add the link to the PR once created.

\section*{Author contributions}
AS conceptualized the paper and proposed the ersatz measure. AP ran the simulations and lead the work. PR wrote the presentation of the classic discrepancy measures, their properties and analysed the results in terms of discrepancy. AP and AS wrote the rest of the paper. All authors revised the final version.

\printbibliography

\end{document}

\pagenumbering{arabic}
\date{}
\title{Discrepancy measures for sensitivity analysis \\ \vspace{4mm} \large{Supplementary Materials}}

\author[1]{Arnald Puy\thanks{Corresponding author}}
\author[2]{Pamphile T. Roy}
\author[3]{Andrea Saltelli}

\affil[1]{\footnotesize{\textit{School of Geography, Earth and Environmental Sciences, University of Birmingham, Birmingham B15 2TT, United Kingdom. E-mail: \href{mailto:a.puy@bham.ac.uk}{a.puy@bham.ac.uk}}}}

\affil[2]{\footnotesize{\textit{Quansight, Vienna, Austria.}}}

\affil[3]{\footnotesize{\textit{Barcelona School of Management, Pompeu Fabra University, Carrer de Balmes 132, 08008 Barcelona, Spain.}}}

%\linenumbers

\maketitle

\tableofcontents

\newpage

\beginsupplement

\section{Figures}

\begin{figure}[!ht]
\centering
\includegraphics[keepaspectratio]{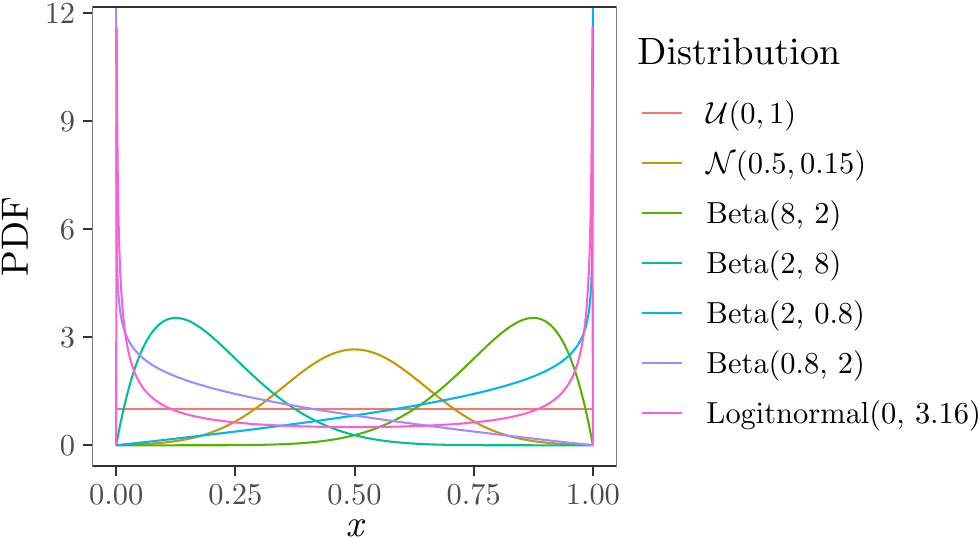}
\caption{Distributions used in the sensitivity analysis.}
\label{fig:SM_plot_distributions}
\end{figure}

\begin{figure}[!ht]
\centering
\includegraphics[keepaspectratio]{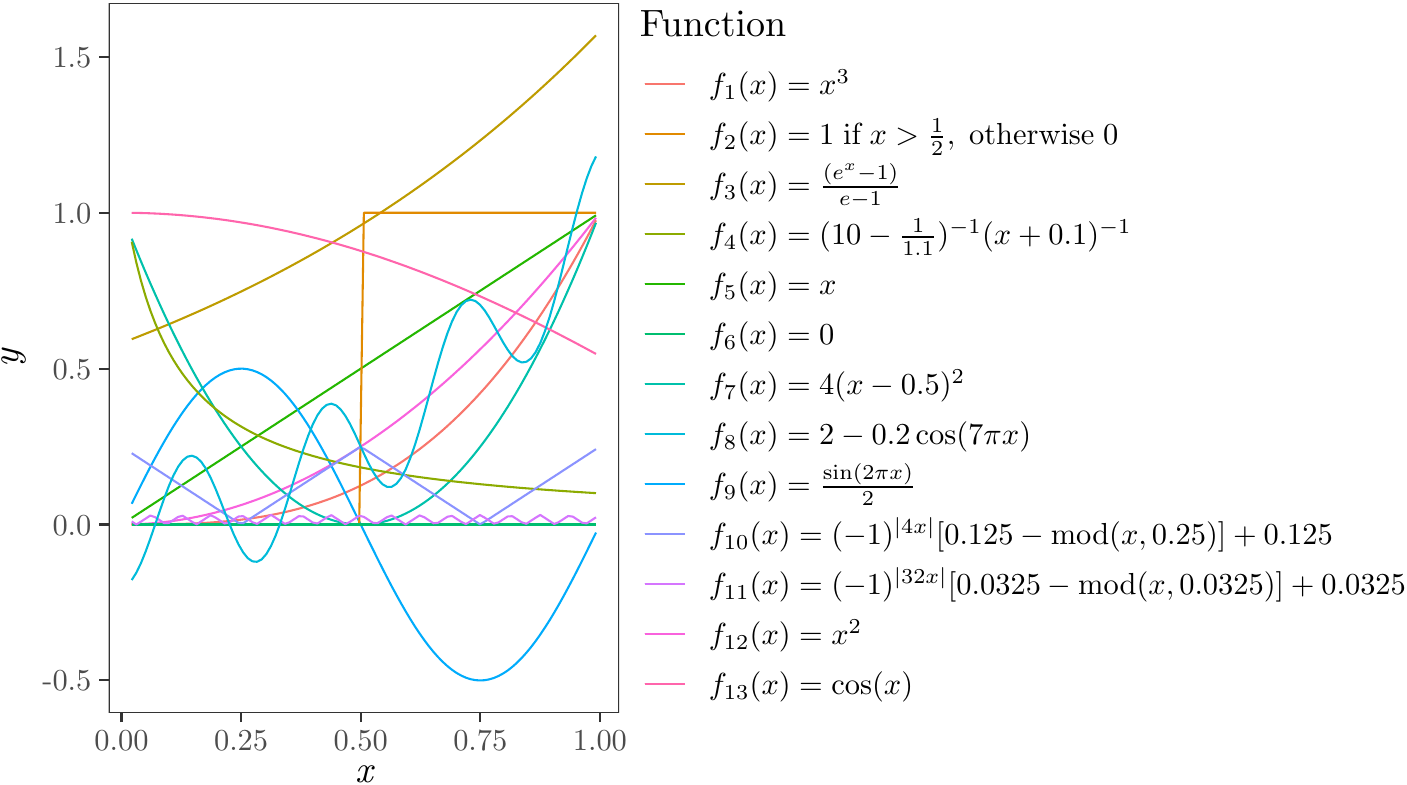}
\caption{Univariate functions included in the meta-model, based on Becker's meta-function~\cite{becker2020}.}
\label{fig:SM_plot_metafunction}
\end{figure}

\begin{figure}[ht]
\centering
\includegraphics[keepaspectratio]{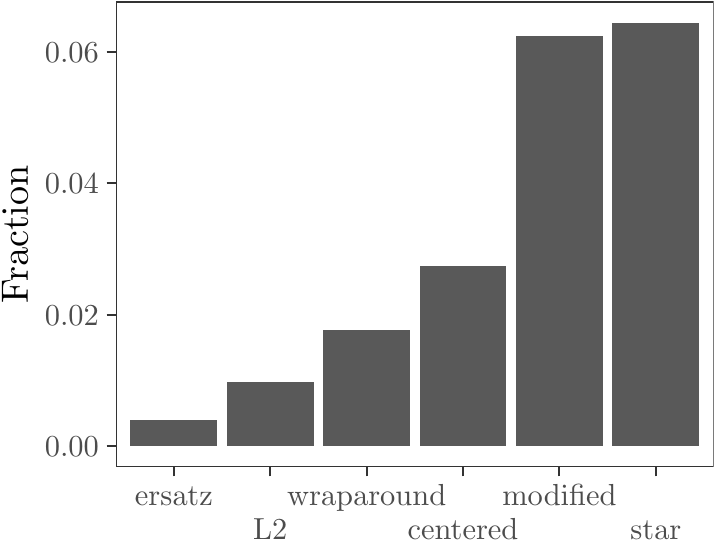}
\caption{Fraction of simulations that yielded $r<0$. The symmetric discrepancy measure is not plotted because all its simulations yielded $r>0$.}
\label{fig:SM_negative}
\end{figure}

\foreach \x in {1,...,10} { 
\begin{figure}[ht]
\centering
\includegraphics[keepaspectratio]{plot_negative-\x}
\caption{Simulations where at least one discrepancy measure yield negative $r$ values. The simulation ID is shown in the facet label.}
\label{fig:SM_negative}
\end{figure}
}

\clearpage

\printbibliography